# Infrared Imaging of the Nanometer-Thick Accumulation Layer in Organic Field-Effect Transistors

Z.Q. Li[1*], G.M. Wang[2], N. Sai[1], D. Moses[2], M.C. Martin[3], M. Di Ventra[1], A. J. Heeger[2] and D.N. Basov[1]

[1] Department of Physics, University of California, San Diego, La Jolla, California 92093, USA

[2] Institute for Polymers and Organic Solids and Mitsubishi Chemical Center for Advanced Materials, University of California, Santa Barbara, Santa Barbara, California 93106, USA

[3] Advanced Light Source Division, Lawrence Berkeley National Laboratory, Berkeley, California 94720, USA

*e-mail: zhiqiang@physics.ucsd.edu

**We report on infrared (IR) spectro-microscopy of the electronic excitations in nanometer-thick accumulation layers in FET devices based on poly(3-hexylthiophene). IR data allows us to explore the charge injection landscape and uncovers the critical role of the gate insulator in defining relevant length scales. This work demonstrates the unique potential of IR spectroscopy for the investigation of physical phenomena at the nanoscale occurring at the semiconductor-insulator interface in FET devices.**

The field-effect transistor (FET) is a benchmark system for exploring the properties of a broad variety of materials as well as for exploiting their novel functionalities. Fundamentally, the electrostatic modulation of carrier density using the FET principle occurs at nanometer scales since the enhanced density of injected charges extends over only few nanometers within the active material.[1] The bottom contact FET devices (schematics in Fig. 1) are particularly well suited for the studies of electrostatic doping of macroscopic samples of novel low-dimensional nanoscale systems[2] such as films of polymer chains,[3] nanotubes,[4] molecules[5] and possibly even DNA bundles,[6] all of which can be easily deposited atop of patterned electrodes. So far, experimental studies of the above FET structures have been primarily limited to transport measurements. New insights into the dynamical properties of the injected carriers are expected from spectroscopic characterization of the electronic excitations in the accumulation layer. However, this is a challenging task given the fact that these layers are exceptionally thin, in the nanometer range. In this letter, we demonstrate the capability of infrared (IR) spectroscopy to explore the electronic excitations in nanometer-thick accumulation layers in bottom-contact FET devices focusing on charge injection in poly(3-hexylthiophene) (P3HT) thin films. We have developed a platform that enables both IR spectroscopic and transport investigations of charge injection in the same device: an essential experimental step towards consistent analysis of transport and spectroscopic data. Earlier IR studies of charge injection in organic thin films have been reported only in metal-insulator-semiconductor structures, which are distinct from bottom-contact FET devices typically employed for transport studies.[7-9]

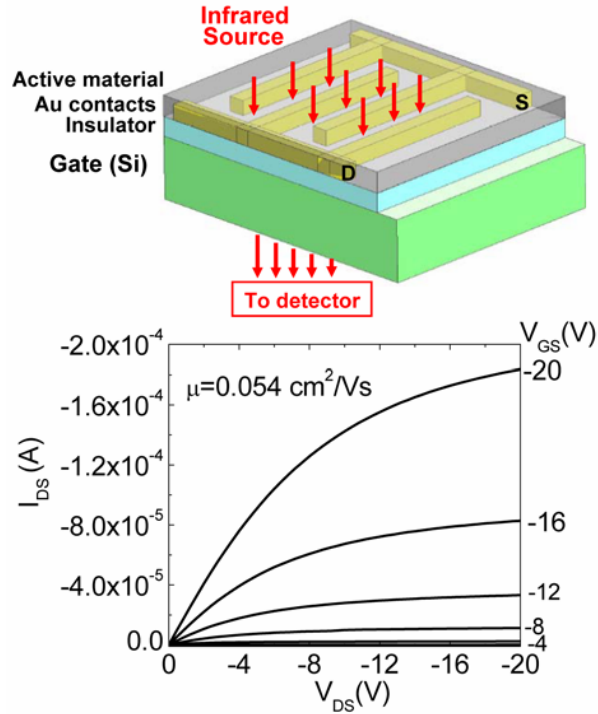

**Figure 1.** Top panel: schematic of a FET device in the bottom-contact geometry for infrared characterization of charge injection. The active material in our FET devices is P3HT. Bottom panel: the I-V curve of a representative $TiO_2$-based FET.

We studied bottom-contact organic FET devices (Fig.1) based on P3HT, a semiconducting polymer with exceptionally high mobility.[10] Devices employing a high dielectric constant (κ) insulator $TiO_2$[11] as well as $SiO_2$[12] were investigated. The goal of using high-κ insulator is to increase the injected carrier density compared to $SiO_2$-based devices.[11] For either type of transistors we have succeeded in probing the electronic excitations in the nanometer thick accumulation layer under applied fields exceeding $10^7$ V/cm: *a regime that has never been explored previously in spectroscopic studies*. An analysis of the oscillator strength of the spectroscopic signatures of charge injection allowed us to quantify the density of the injected carriers and examine its evolution with applied voltages. Using IR microscopy we were able to monitor the spatial dependence of the injected charges in the active area of the device. Our results for the high-κ devices show significant departures from the behavior expected for an "ideal" FET[13] in which the charge density increases linearly with voltages and is uniform in the channel. This study uncovers the unique potential of IR spectroscopy for investigating the dynamical properties of the electronic excitations in FET structures.

Large area FET devices[14] (>1 cm$^2$) with gate insulator deposited on n-Si were investigated in this work. We employed two types of gate insulators: 200 nm thick $SiO_2$ and $SiO_2$(6 nm)/$TiO_2$(180 nm) bilayer; we will refer to the latter devices as "$TiO_2$-based". The transport mobility of P3HT in our $SiO_2$-based transistors[12] is 0.18 cm$^2$V$^{-1}$s$^{-1}$, whereas that in $TiO_2$-based FETs[11] is 0.05 cm$^2$V$^{-1}$s$^{-1}$. The I-V curve for a typical $TiO_2$-based FET is shown in the bottom panel of Fig. 1. In these FET devices, source and drain Au electrodes (with a spacing of 50-200 μm) were patterned on insulating oxides followed by the deposition of a 4 – 6 nm-thick P3HT film. Fig.1 shows a cross-section of the devices whereas Fig.4d depicts a top view photograph of an actual device. The breakdown voltage of $TiO_2$-based devices is about -35~ -45 V and that of $SiO_2$ -based FETs exceeds -100V. In bottom-contact FET devices, an applied gate voltage induces an accumulation layer[3,15] in P3HT that forms the p-type conducting channel between the Au electrodes. This channel is not obscured by any other interfaces from above and is therefore well suited for the spectroscopic studies[14] of the accumulation layer in the polymer film from far-IR to near-IR with the latter cut-off imposed by the band gap of Si substrate. To examine the length scales associated with charge injection we fabricated devices with a "V-shape" electrode pattern (Fig, 4b). We studied changes of transmission as a function of applied gate voltage $V_{GS}$ normalized by the transmission at $V_{GS}$=0: $T(\omega,V_{GS})/T(\omega,V_{GS}=0)$. The source and drain electrodes were held at the same potential in most measurements. All the data reported here were recorded at room temperature with a spectral resolution of 4 cm$^{-1}$.

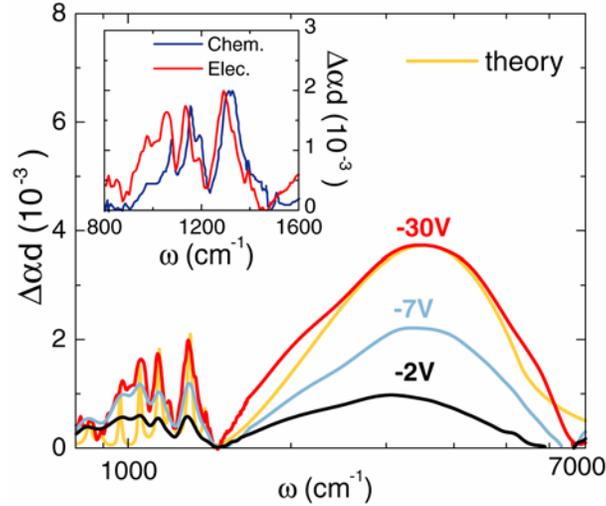

**Figure 2.** The voltage-induced absorption spectra $\Delta\alpha d$ for the P3HT layer under applied gate voltages $V_{GS}$ in a TiO$_2$-based device. The green curve is a theoretical modeling of the experimental spectrum as detailed in Supporting Information. **Inset:** $\Delta\alpha d$ spectrum for a representative TiO$_2$ based device at $V_{GS}$=-30V along with the data for chemically doped P3HT.[18] The latter is the difference spectrum between the absorption of chemically doped P3HT with 1 mol % PF$_6^-$ and that of pure P3HT[18] scaled by a factor of 4*10$^5$. All spectra uncover spectroscopic fingerprints of electrostatic doping: IRAV modes in 1,000-1,500 cm$^{-1}$ range and a polaron band at 3,500 cm$^{-1}$. The oscillator strength of both the polaron band and the IRAV modes increases with gate voltage. The noise of the $\Delta\alpha d$ spectra is less than 10$^{-4}$.

The absorption spectra $\Delta\alpha d$ = 1-T($V_{GS}$)/T(0V) of TiO$_2$-based devices are displayed in Fig. 2. Here $\Delta\alpha$ is the change of the absorption coefficient of P3HT with applied voltage and d is the thickness of the accumulation layer. These spectra show two voltage-induced features: i) sharp resonances in the 1,000 -- 1,500 cm$^{-1}$ region, and ii) a broad band centered around 3,500 cm$^{-1}$. A gradual development of these features with increasing gate voltage $V_{GS}$ suggests that they are intimately related to the formation of charge accumulation layer in P3HT.[16] This assignment is supported by earlier reports of similar changes of optical properties produced by photoexcitation,[17-19] chemical doping[17,18] or electrostatic charge doping achieved by placing a P3HT film between semi-transparent electrodes.[7-9] Sharp resonances in the 1,000-1,500 cm$^{-1}$ range result from the IR active vibrational modes (IRAVs); i.e Raman modes made IR active by distortions of the polymer backbone caused by the self-localized charges.[17] The frequencies of the IRAV modes are in excellent agreement with the vibrational resonances found in chemically doped P3HT (inset of Fig. 2). The broad absorption band centered around 3,500 cm$^{-1}$ is usually ascribed to a midgap state of polaron or bipolaron associated with the local relaxation of the lattice around the doped charge.[17] Whether this absorption is due to polaron or bipolaron is still under investigation.[8,18] We'll refer to this broad absorption band as polaron for simplicity. Both the IRAV modes and the polaron band can be quantitatively described by the amplitude mode model[20,21] of charge excitations in conjugated polymers. A theoretical fit of the experimental results based on this model is shown in Fig. 2 (see Supporting Information for details). Similar features due to IRAV modes and polaron are also observed in SiO$_2$-based devices as will be discussed in details in an upcoming publication.

With the key spectroscopic signatures of charge injection in P3HT established for our open channel devices (Fig. 2) we now turn to the analysis of their oscillator strength. It is instructive to define the effective spectral weight as $N_{eff} = \int (\Delta \alpha d) d\omega$, which is proportional to the 2D density of the injected charges responsible for the absorption structure in our data. Notably, the polaron band and IRAV modes are well separated from each other in the spectra in Fig.2 and therefore the oscillator strength of these two structures can be quantified by properly choosing the integration cut-offs. Fig. 3 displays the spectral weight of the polaron band $N_{eff}^{P}$ (integrated from 1,450 cm$^{-1}$ to 6,000 cm$^{-1}$) and that of the IRAV modes $N_{eff}^{IRAV}$ (integrated from 900 cm$^{-1}$ to 1,450

cm$^{-1}$) plotted as a function of V$_{GS}$. A gradual growth of both $N_{eff}^{P}$ and $N_{eff}^{IRAV}$ with the increase of V$_{GS}$ is observed. The simple capacitive model of an FET device predicts the linear dependence between the charge density and the bias voltage with the slope determined solely by the dielectric constant of the gate insulator $\kappa$ and the thickness of the insulator L:

$$N_{2D} = \frac{\kappa \varepsilon_0}{eL} V_{GS}. \qquad (1)$$

Devices based on TiO$_2$ reveal a linear voltage dependence of $N_{eff}(V_{GS})$ at small V$_{GS}$ with an obvious trend to saturation at higher biases.

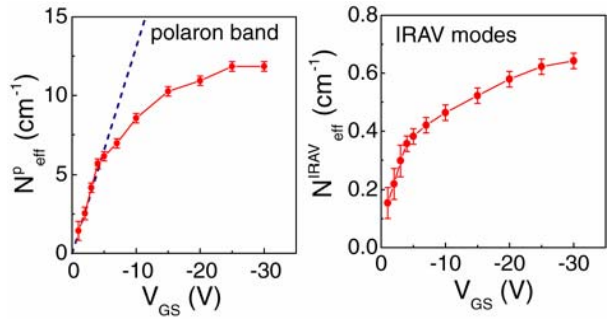

**Figure 3.** Evolution of the spectral weight of the polaron band $N_{eff}^{P}(V_{GS})$ (left panel) and of the IRAV modes $N_{eff}^{IRAV}(V_{GS})$ (right panel) with gate voltage V$_{GS}$ in a TiO$_2$-based device. The dashed line in the left panel represents the linear V$_{GS}$-dependence of $N_{eff}^{P}$ expected from a capacitive model.

Below we show that valuable insights into the accumulation layer characteristics are provided by a survey of the spatial distribution of charge density in the FET devices using IR microscopy. We carried out microscopic study of the excitations associated with the electrostatically doped charges in P3HT using the infrared beamlines at the Advanced Light Source (ALS) facility. With the focused beam of an IR microscope we were able to record spectra similar to those displayed in Fig.2 from areas as small as 50-100 μm in diameter. The IR beam was scanned in between the "V" shaped electrodes (Fig. 4b) or in the corner of the electrodes in Fig. 4d with simultaneous monitoring of the voltage-induced changes in the spectra. The frequency dependence of absorption spectra did not change appreciably throughout the entire device. We therefore focus on the spatial dependence of the integrated weight of both IRAVs and of the polaron band. In Fig.4a we plot the spectral weight of the IRAV modes of P3HT as a function of separation between the V-shaped electrodes $l$ normalized by data at $l$=0: $N_{eff}^{IRAV}(l)/N_{eff}^{IRAV}(l=0)$. $TiO_2$-based FETs reveal a gradual decay of the injected charge density away from the electrodes that vanishes at length scales of about 500 μm. The decay of the injected carrier density in FETs with $TiO_2$ gate insulator is also evident in the 2D charge density profile $N_{eff}^{IRAV}(x,y)$ shown in Fig. 4c. Here $N_{eff}^{IRAV}(x,y)$ vanishes at distances beyond 500-600 μm away from the electrodes in accord with the data for V-shaped structures. On the contrary, FETs structures with $SiO_2$ gate insulator show no change of carrier density at least up to 1.6 mm away from the contacts limited only by the physical dimensions of our devices. This latter result verifies that in $SiO_2$-based FETs a uniform equipotential layer is formed consistent with the notion of an "ideal" field effect transistor.[13] The charge injection landscape was also explored by imaging the polaron absorption with a spatial resolution of 3 μm in a set-up based on synchrotron source at the ALS. These fine resolution results are identical to those inferred from IRAV modes: a gradual decrease of $N_{eff}^{P}$ in $TiO_2$-based devices and no measurable decrease of $N_{eff}^{P}$ in $SiO_2$-based structures. Neither result significantly depends on the biasing voltage. To the best of our knowledge, this is the first spatially-resolved IR imaging of the injected charges in FETs.

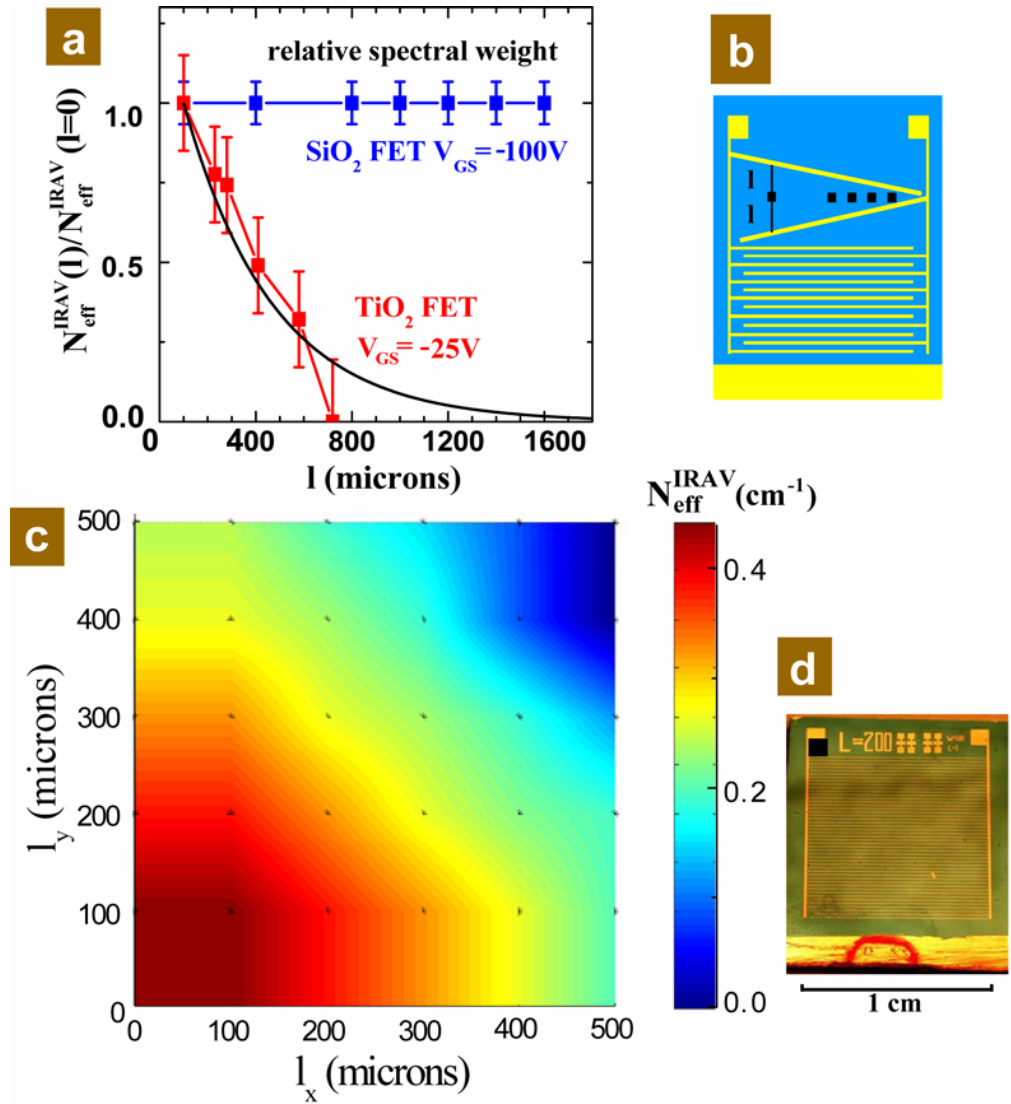

**Figure 4.** Infrared imaging of the charge injection landscape in several representative FETs. **a**, the spectral weight of IRAV modes of P3HT as a function of separation between the V-shaped electrodes *l* normalized by data at *l*=0: $N_{eff}^{IRAV}(l)/N_{eff}^{IRAV}(l=0)$, indicated by the black squares in **b** for devices with "V" shape electrodes. The black curve in **a** shows a fit of N(l)/N(l=0) with an exponential form.[24] **c**, the 2D charge profile $N_{eff}^{IRAV}$ ($l_x$,$l_y$) in the P3HT layer of an FET with TiO$_2$ gate insulator under -25V. The mapping region is the up-left corner of the electrode as schematically shown in **d** by the black square. The charge injection landscape inferred from the polaron absorption is identical to **a** and **c** generated via monitoring of the IRAV modes. The length scale of the charge injection process in FETs with TiO$_2$ gate insulator is several hundred microns, whereas the injected charges form a uniform layer in FETs with SiO$_2$ gate insulator and the charge injection length scale is 1.6mm or even longer.

The saturation behavior (Fig. 3) of the spectral weight of *localized excitations*, i.e., $N_{eff}^{P}(V_{GS})$ and $N_{eff}^{IRAV}(V_{GS})$, may be indicative of the precursor to the insulator-to-metal transition in conjugated polymers[22,23] at high carrier density. In the vicinity of the metallic state, the injected charges acquire more extended character in direct competition with the formation of IRAV or polaronic resonances. However, this intriguing interpretation has to be critically examined since several other factors can in principle mimic the saturated or nonlinear voltage dependence of the localized modes, such as the high leakage currents exceeding hundred μA at high voltages observed in $TiO_2$-based FETs. A gradual decay of charge density in the accumulation layer in $TiO_2$-based FETs revealed by IR imaging experiments probably originates from the imperfections of the gate dielectric/polymer interface as well as high leakage currents. Naturally, charges that are either trapped at the insulator/polymer interface or leak through the insulator do not contribute to the oscillator strength of the features in the absorption spectra. Therefore injected charges are registered in our imaging studies only after they exhaust all potential leakage paths through the insulator and traps at the polymer/insulator interface. The length scale limiting propagation of charges away from the injection contacts naturally follows as a result of competition between the channel resistance and the leakage resistance of the gate insulator.[24] These IR microscopy results indicate that the saturation of the oscillator strength of localized excitations at high $V_{GS}$ biases is extrinsic and originates from the limitations of $TiO_2$ gate insulator.

In summary, IR spectroscopy investigations of the electronic excitations in a nanometer thick accumulation layer, i.e., IRAV modes and polarons, have been carried out in P3HT thin film FET devices. We show that the unconventional behavior in $TiO_2$-based FETs is due to the limitations of the $TiO_2$ gate insulator. Our work has demonstrated that IR spectroscopy is a unique technique for the study of charge injection in macroscopic samples of nanometer-scale materials in bottom-contact FET devices. Instrumental innovations reported here uncover the potential of IR spectro-microscopy for the investigation of both IRAV modes and polarons, which are of fundamental importance for the understanding of charge transport in other materials as well, such as DNA.[6,25,26]

We thank M. Fogler for useful discussions. Work at UCSD is supported by NSF and PRF. Work at UCSB is partially supported by the NSF under DMR0099843. The Advanced Light Source is supported by the Director, Office of Science, Office of Basic Energy Sciences, Materials Sciences Division, of the U.S. Department of Energy under Contract No. DE-AC03-76SF00098 at Lawrence Berkeley National Laboratory.

**Supporting Information**

Theoretical description of the field-induced absorption spectrum of P3HT.

We have fitted the field-induced absorption spectrum of P3HT using the amplitude mode model[20,21] of charge excitations in conjugated polymers. This theory characterizes the vibrational excitations as multiple normal phonon modes coupled to the same electronic dimerization pattern and it has been sucessfully used in explaining the Raman modes and charge-induced Infrared active modes. In the generalized theory[21] for systems that do not satisfy the adiabatic approximation, i.e., all phonon frequencies are much smaller than the electronic band gap, the frequency dependent electrical conductivity is given by

$$\sigma(\omega) = \frac{\omega_P^2}{4\pi i \omega}\left\{f(\omega)\frac{1+D_0(\omega)[1-\alpha]}{1+D_0(\omega)[1+c(\omega)-\alpha]}-1\right\},$$

where $D_0(\omega) = \sum \frac{\lambda_n}{\lambda}\frac{\omega_{n0}^2}{\omega^2-\omega_{n0}^2}$ is the phonon response function for the vibrational modes, $\omega_{n0}$ and $\lambda_n$ are, respectively, the bare phonon frequencies and electron-phonon coupling parameter for the $n$th mode. Other parameters in this model are $f(\omega) = \frac{E_r^2}{\omega^2 y}\arctan(1/y)$ when $\omega$<E$_r$, and $f(\omega) = \frac{E_r^2}{2\omega^2 y}[\ln\frac{1-y}{1+y}+i\pi]$ when $\omega$>E$_r$, where $y = \sqrt{|1-E_r^2/\omega^2|}$, E$_r$ is the polaron relaxation energy, $c(\omega) = \frac{\lambda \omega^2 f(\omega)}{E_r^2}$, $\alpha$ is the pinning parameter associated with the field-induced polarons and $\lambda = \sum \lambda_n$ is the dimensionless coupling constant. Nine bare phonon frequencies are used to fit the experimental spectrum in Fig. 2, all of which are in good agreement with the phonon frequencies extracted from experiment in Ref. 20. The fitting parameters used for the theoretical spectrum in Fig. 2 are E$_r$=3,380 cm$^{-1}$, $\lambda$=0.2 and $\alpha$=0.09. The good agreement between the experimental spectrum and that from theoretical modeling corroborates the microscopic origins of the observed field-induced absorption spectrum in Fig. 2.

References

(1) Ando, T.; Fowler, A. B.; Stern, F. Rev. Mod. Phys. 1982, 54, 437.

(2) Di Ventra, M.; Evoy, S.; Heflin, R. eds. *Introduction to Nanoscale Science and Technology*; Kluwer Academic Publishers: 2004.

(3) Dimitrakopoulos, C.D.; Malenfant, P.R.L. *Adv. Mater.* 2002, *14*, 99.

(4) Snow, E. S.; Novak, J. P.; Campbell, P. M.; Park, D. *Appl. Phys. Lett.* 2003, *82*, 2145.

(5) Joachim, C.; Gimzewski, J. K.; Aviram, A. *Nature* 2000, *408*, 541.

(6) Di Ventra, M.; Zwolak, M. *DNA electronics* in Encyclopedia of Nanoscience and Nanotechnolog, Vol. 2, p. 475. H. S. Nalwa ed. American Scientific Publishers, 2004.

(7) Sirringhaus, H. *et al. Nature* 1999, 401, 685.

(8) Brown, P.J.; Sirringhaus, H.; Harrison, M.; Shkunov, M.; Friend, R. H. *Phys. Rev. B* 2001, 63, 125204.

(9) Ziemelis, K. E. *et al. Phys. Rev. Lett.* 1991, 66, 2231.

(10) Hamadani, B. H.; Natelson, D. *Appl. Phys. Lett.* 2004, 84, 443.

(11) Wang, G. *et al. J. Appl. Phys.* 2004, 95, 316.

(12) Wang, G. M.; Swensen, J.; Moses, D; Heeger, A. J. *J. Appl. Phys.* 2003, 93, 6137.

(13) Sze, S.M. *Physics of Semiconductor Devices*. 2$^{nd}$ ed. Wiley: New York, 1981.

(14) Li, Z.Q. *et al. Appl. Phys. Lett.* 2005, 86, 223506.

(15) Li, T.; Balk, J.W.; Ruden, P.P.; Campbell, I.H.; Smith, D.L. *J. Appl. Phys.* 2002, 91, 4312.

(16) Due to the imperfections of the $TiO_2$/Si boundary as well as of the surface phonon scattering (Fischetti, M.V.; Neumayer, D.A.; Cartier, E.A. *J. Appl. Phys.* 2001, 90, 4587) originating from the large polarizability of high-k insulator $TiO_2$, absorption associated with the accumulation layer in n-Si is spread out over a broad frequency range and has a negligible contribution to the absorption spectra in the far-IR[14] and mid-IR. Therefore, the Dαd spectra displayed in Fig. 2 can be attributed to the voltage-induced carriers in P3HT.

(17) Heeger, A. J.; Kivelson, S.; Schrieffer, J. R.; Su, W.-P. *Rev. Mod. Phys.* 1988, 60, 781.

(18) Kim, Y. H.; Spiegel, D.; Hotta, S.; Heeger, A. J. *Phys. Rev. B* 1988, 38, 5490.

(19) Österbacka, R.; Jiang, X. M.; An, C. P.; Horovitz, B.; Vardeny, Z.V. *Phys. Rev. Lett.* 2002, 88, 226401.

(20) Horovitz, B. *Solid State Commun.* 1982, 41, 729.

(21) Horovitz, B.; Österbacka, R.; Vardeny, Z.V. *Synth. Metals* 2004, 141, 179.

(22) Kohlman, R.S.; Epstein, A.J. *Insulator-Metal Transition and Inhomogeneous Metallic State in Conducting Polymers* in *Handbook of Conducting Polymers*; Skotheim, T. A.; Elsenbaumer, R. L.; Reynolds, J. R. eds. Marcel Dekker: New York, 1998.

(23) Kivelson, S.; Heeger, A. J. Phys. Rev. Lett. 1985, 55, 308.

(24) Assuming there exists a large number of uniformly distributed leakage channels through the insulator along the 1D conducting channel of a FET, one can show that the charge density N decays exponentially as a function of the distance l from the injection source, i.e., $N(l) = N(l=0)\exp[-\frac{R_1}{R_{eq}}\frac{l}{\rho}]$, where $R_{eq} = \frac{1}{2}(R_1 + \sqrt{R_1^2 + 4R_1R_2})$, $R_1$ is the resistance of the conducting channel between leakage paths, $R_2$ is the resistance of the leakage channel and $\rho$ is the number of leakage channels per unit length along the conducting channel. (See, e.g., E.M. Purcell, *Electricity and Magnetism*, Chapter 4, Berkeley Physics Course Vol. 2. McGraw-Hill,

New York, 1965). The black curve in Fig. 4a shows a fit of N(l)/N(l=0) using the above exponential form with $\frac{R_1}{\rho R_{eq}}$ =0.0027. (For a 2D channel N(l) may have a different decay form.)

(25) Endres, R. G.; Cox, D. L.; Singh, R. R. P. Rev. Mod. Phys. 2004, 76, 195.

(26) Omerzu, A. *et al.*, Phys. Rev. Lett. 2004, 93, 218101.